\documentclass[12pt]{article}
\usepackage[latin1]{inputenc}
\usepackage[dvips]{graphicx}
\usepackage{graphicx}
\newcommand{\drm}{{\rm d}}

\newcommand{\text}{\rm}

\newcommand{\ug}{ \; = \; }

\newcommand{\bb}{\begin{equation}}
\newcommand{\ee}{\end{equation}}
\newcommand{\bega}{\begin{eqnarray}}
\newcommand{\ega}{\end{eqnarray}}
\newcommand{\begae}{\begin{eqnarray*}}
\newcommand{\egae}{\end{eqnarray*}}

\newcommand{\h}{\hspace*{4ex}}
\newcommand{\dis}{\displaystyle}

\newcommand{\om}{\omega}

\newcommand{\cent}{\centerline}
\newcommand{\vs}{\vspace*}

\begin{document}

\baselineskip 0.8cm

\begin{center}

{\Large {\bf Analytical expressions for
the longitudinal evolution of nondiffracting pulses
truncated by finite apertures.}}

\end{center}


\vs{3mm}

\cent{ Michel Zamboni-Rached, }

\vs{0.1 cm}

\centerline{{\em D.M.O., Faculty of Electrical Engineering,
UNICAMP, Campinas, SP, Brasil.}}

\vs{0.5 cm}

\

\begin{abstract}
In this paper, starting from some general and plausible
assumptions based on geometrical optics and on a common feature of
the truncated Bessel beams, a heuristic derivation is presented of
very simple {\em analytical} expressions, capable of describing
the longitudinal (on-axis) evolution of axially-symmetric
nondiffracting pulses when truncated by finite apertures. We apply
our analytical formulation to several situations involving
subluminal, luminal or superluminal localized pulses, and compare
the results with those obtained by numerical simulations of the
Rayleigh-Sommerfeld diffraction integrals. The results are in
excellent agreement. The present approach can be rather useful,
because it yields, in general, closed-form expressions,
avoiding the need of time-consuming numerical simulations; and
also because such expressions provide a powerful tool for
exploring several important properties of the truncated localized
pulses, as their depth of fields, the longitudinal pulse behavior,
the decaying rates, and so on.
\end{abstract}

\ .\hfill\break


\section{Introduction}

\h Ideal nondiffracting pulses (INP) are infinite energy solutions
of the ordinary linear wave equation, capable of maintaining their
spatial shapes indefinitely (sometimes with just small local
variations) while propagating[1-9]. When these ideal solutions are
to adapted to real situations and applications, they must be
spatially truncated by a finite aperture (i.e., generated by a
finite aperture), getting transformed into finite energy
solutions, with finite field depths, even if these field-depths
are very large when compared to those of ordinary pulses.

\h When we truncate an INP, the resulting wave field cannot be
obtained, in general, in analytical form. In this case one has to
resort to the diffration theory and perform numerical simulations
of the diffraction integrals, such as that, well known, of
Rayleigh-Sommerfeld[10].

\h Indeed, one can get very important pieces of information about
a truncated nondiffracting pulse (TNP) by performimg numerical
simulations of its longitudinal evolution[11-14], especially when
the pulse is axially symmetric.

\h In this paper we shall show that, by using some general and
plausible assumptions, based on geometrical optics and on a common
feature of truncated Bessel beams, a heuristic derivation of
simple analytical expressions is possible, capable of furnishing
the longitudinal (on-axis) evolution of the TNPs. It is
interesting to notice that this approach depends only on the
spectral structure of the relevant INP.

\h We compare the results of our analytical method, when applied
to several different situations involving subluminal, luminal or
superluminal TNPs, with the results obtained from the usual numerical
simulation of the Rayleigh-Sommerfeld integrals: the results
are in an excellent agreement.

\h This method, due to its extreme simplicity and analytical
character, can be a powerful tool for exploring several important
properties of the TNPs, as their depth of fields, the longitudinal
pulse behavior, the decaying rates, etc.;  for revealing the effects of
the spectral parameters on the pulses evolution; and also for
comparing the ``effectiveness" of the different kinds of TNPs, as the
subluminal, luminal and superluminal ones. Without this method, all
those results could be reached (in each particular situation) only
by performing several time-consuming numerical simulations.

\section{Heuristic approach for describing the on-axis evolution of TNPs}

\h Let us begin this Section by making some comments about the
truncated Bessel beams and about some approximations, which will be used
below for developing the method here proposed.

\subsection{Some observations and approximations about truncated Bessel beams}

\h An ideal (infinite energy) Bessel beam (IBb) is given by[15]

\bb \Psi_{\rm IBb}(\rho,z,t) \ug J_0(k_{\rho}\rho)e^{i\beta
z}e^{-i\om t} \;\; ,  \label{ibb}
 \ee

 where $\rho^2=x^2+y^2$ is the transverse coordinate, \ $k_{\rho}^2=\om^2/c^2 -
 \beta^2$ is the transverse wavenumber, \ $\beta$ is the
 longitudinal wavenumber, and $\om$ is the angular frequency.

 \h An important parameter[16-19] of an IBb is its axicon
 angle $\theta$, where $\om = c\beta/\cos\theta$.

 \h When a Bessel beam is truncated by a finite aperture of radius
 $R$, Eq.(\ref{ibb}) cannot be used to describe the resulting beam
 in the whole space. However, if the size of the aperture is big
 enough to contain several bright rings of the ideal incident
 Bessel beam, i.e., if $R>>1/k_{\rho}$, we can use geometrical
 optics to get some characteristics of the truncated Bessel
 beam (TBb) evolution.

 \h In this case (see Fig.1), we can say that, in the spatial region
 localized inside the cone oriented along the $z$-axis, with apex at
 $z=Z=R/\tan\theta$ and base given by the circular aperture, the
 resulting TBb can be approximately described by Eq.(\ref{ibb}).

 \

\begin{figure}[!h]
\begin{center}
 \scalebox{3}{\includegraphics{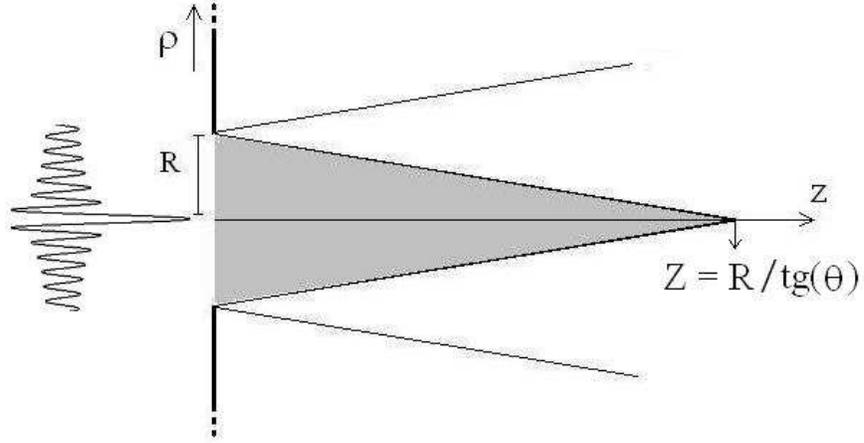}}
\end{center}
\caption{A typical Bessel beam truncated by a finite aperture.}
\label{Fig1}
\end{figure}

\

 \h However, when using geometrical optics, after the
 distance $Z=R/\tan\theta$
 the on-axis amplitude of the TBb becomes approximately zero (see
 Fig.1). The distance $Z$ is called the depth-field of the TBb.

 \h Keeping the above observations in mind, one may affirm that, since
 $R>>1/k_{\rho}$, the on-axis behavior of a TBb can be
 approximately described by

 \bb
 \Psi_{\rm TBb}(\rho=0,z,t) \simeq \left\{\begin{array}{clcr}
e^{i\beta z}e^{-i\om t} \;\;\;  &{\rm for} \;\;\; z \leq
\frac{R}{\tan\theta}\\
\\
 0 \;\;\;  &{\rm for} \;\;\; z > \frac{R}{\tan\theta}
\end{array}\right.
\ee

which can be compactly written as

\bb
 \Psi_{\rm TBb}(\rho=0,z,t) \, \simeq \, e^{i\beta z}\,e^{-i\om
 t}\,\left[H(z) - H\left(z-\frac{R}{\tan\theta}\right)\right] \label{TBb}
\ee

where $H(.)$ is the Heaviside step function. {\it The equation
above is the starting point for our heuristic method
describing the on-axis TNP's behavior}.

\h According to the approximation Eq.(\ref{TBb}), the on-axis field
intensity of a TBb is a rectangular function with unitary value,
till $z=Z$.

\h On the other hand, when a numerical simulation of the
diffraction integrals is performed[8,15], one can observe that the
TBb presents some on-axis field oscillations around the unitary
value,
before suffering an abrupt decay after $z=Z$. \  
Such oscillations cannot be predict by
geometrical optics, and arise only due to the abrupt truncation
made by the aperture. However, it is important to stress here that,
despite the fact that those oscillations are not predicted by Eq.(\ref{TBb}),
such an error {\it is not present}, in general, in the case of
our description of truncated localized pulses. One can understand
this by noticing that, since nondiffracting pulses are constructed
through Bessel beams superpositions, those oscillations,
originating from each TBb, suffer a destructive interference.

\h With all this in mind, we are ready to develop our method.

\subsection{The heuristic approach}

\h It is well known[1-9] that axially-symmetric ideal
nondiffracting pulses (INP) can be made by zero order Bessel beam
superpositions,

\bb \Psi(\rho,z,t) \ug
\int_{0}^{\infty}\,d\om\,\int_{-\om/c}^{\om/c}\,d\beta\,
\overline{S}(\om,\beta) J_0\left(\rho\sqrt{\frac{\om^2}{c^2} -
\beta^2}\right)e^{i\beta z}e^{-i\om t} \label{bbsup} \ee

{\it\bf provided that} the spectral function
$\overline{S}(\om,\beta)$ entails a linear relationship of the type

\bb \om \ug V\,\beta + b \label{cond2}\ee

between $\om$ and $\beta$. In this way, by putting
$\overline{S}(\om,\beta)=S(\om)\delta(\om-V\beta-b)$, the general
form of an axially-symmetric INP gets written as

\bb \Psi_{INP}(\rho,z,t) \ug
e^{-ibz/V}\,\int_{\om_{min}}^{\om_{max}}\,d\om\, S(\om)
J_0\left(\rho\sqrt{\left(\frac{1}{c^2} - \frac{1}{V^2}
\right)\om^2 + \frac{2\,b\,\om}{V^2} -
\frac{b^2}{V^2}}\right)e^{i\om\zeta/V} \label{inp} \ee

\

where $\zeta=z-Vt$, and $S(\om)$ is the frequency spectrum.
Obviously, the INP will be subluminal, luminal or supeluminal,
depending on the value of $V$, it being be $<$, $=$ or $>c$. The positive
quantities $\om_{min}$ and $\om_{max}$ are the minimum and maximum
angular frequency allowed for the Bessel beams in the superposition
(\ref{inp}), and their values have to be estimated as follows.


\h Once we have chosen the value of $V$ in (\ref{inp}), the values
of $b$, $\om_{min}$ and $\om_{max}$ are to related between them
in such a way that

\bb \begin{array}{clcr} k_{\rho}^2 \ug \dis{\left(\frac{1}{c^2} -
\frac{1}{V^2} \right)\om^2 + \frac{2\,b\,\om}{V^2} -
\frac{b^2}{V^2}} \, \geq \, 0\\
\\
\beta \geq 0 \end{array} \label{cond3} \ee

for all positive angular frequency $\om_{min} \leq \om
\leq\om_{max}$ used in the superposition (\ref{inp}).

\h In relation (\ref{cond3}), the condition on $k_{\rho}$ eliminates any
unphysical behaviors of Bessel functions and evanescent waves in
Eq.(\ref{inp}). The second condition ($\beta \geq 0$) eliminates
any backwards-travelling Bessel beams in the same superposition, since we are
considering positive values of the angular frequency only.

\h Taking into account conditions (\ref{cond3}), one has:

\begin{itemize}
    \item For Subluminal ($V<c$) INPs: $b>0$, $\om_{min}=b$ and
    $\om_{max}=cb/(c-V)$
    \item For Luminal ($V=c$) INPs: $b>0$, $\om_{min}=b$ and
    $\om_{max}=\infty$
    \item For Superluminal ($V>c$) INPs: $b \geq 0$, $\om_{min}=b$ and
   $\om_{max}=\infty$. Or $b<0$, $\om_{min}=cb/(c-V)$ and $\om_{max}=\infty$

\end{itemize}

\h The INPs provided by Eq.(\ref{inp}) can propagate without
distortion indefinitely, with peak-velocity $V$.

\h Such INPs possess an infinite energy, and so, for real
applications, they must be spatially truncated (i.e., generated by
finite apertures)[11-14], resulting in finite energy solutions,
with a finite depth of field.

\h When such truncation is made, the resulting pulse in general
cannot be obtained in an analytically form, but has to be
numerically calculated from the diffraction theory, by using, for
example, the Rayleigh-Sommerfeld formula[10-14]. That is, once we
have a known INP solution $\Psi_{INP}$, its truncated version
$\Psi_{TNP}$, generated by a finite aperture of radius $R$ on the
plane $z=0$, results given by

\bb \Psi_{TNP}(\rho,z,t) \ug \int_0^{2\pi}\drm\phi'\int_0^{R}
\drm\rho'\rho' \frac{1}{2\pi D}\left\{[\Psi]\frac{(z-z')}{D^2} \,
+ \, [\partial_{ct'}\Psi]\frac{(z-z')}{D}\right\} \label{rs} \ee

the quantities enclosed by the square brackets being evaluated at
the retarded time $ct'=ct-D$. The distance
$D=\sqrt{\left(z-z'\right)^2+\rho^2+\rho'^2-2\rho\rho'\cos\left(\phi-\phi'%
\right)}$ is the separation between the source and observation points.

\h Due to its complexity, Eq.(\ref{rs}) has to be solved
numerically in most cases.

\h Of particular interest is the on-axis behavior of $\Psi_{TNP}$.
Actually, many important information can be extracted from the
evolution of $\Psi_{TNP}(\rho=0,z,t)$, like its depth of field, the
pulse decaying rate, its longitudinal behavior, the effects of the
different spectral parameters on the pulse evolution; and, even more
important, quantity $\Psi_{TNP}(\rho=0,z,t)$ can be used to compare the
performance of different kinds of TNP, as for example, the luminal
and the superluminal ones.

\h On considering axially-symmetric TNP, and making $\rho=0$ in
Eq.(\ref{rs}), we get some simplifications, because the integration
on $\phi'$ can be immediately done: But, even in this case, the
integration on $\rho'$ rarely can be carried on analytically, due to the
complexity of the integrand, and numerical simulations are once more
required.

\h To overcome this problem, let us propose the following heuristic
approach:

\h First, we consider the Bessel beam superposition (\ref{inp}),
which provides us with the INPs.

\h Second, we make the assumption that each Bessel beam, with
frequency $\om$ and axicon angle $\theta(\om)$, entering in
superposition (\ref{inp}), obeys the following condition:

\bb R >> \frac{1}{k_{\rho}}=\frac{c}{\om\sin\theta(\om)} \;,
\label{cond4}\ee

$R$ being the radius of the finite aperture that will be used for
truncating the INP.

\h The above assumption is very plausible, since efficient TNPs are
generate by large apertures.

\h Once condition (\ref{cond4}) is fulfilled by all Bessel beams
in superposition (\ref{inp}), we can use again the geometrical optics
and assume that, after the truncation, the on-axis behavior of each
one of those Bessel beams can be approximated by Eq.(\ref{TBb}).

\h Third, taking into account Eqs.(\ref{TBb}) and (\ref{inp}), we
may conjecture that the on-axis evolution of the {\it truncate}
nondiffracting pulse is approximately given by:

\bb \Psi_{TNP}(\rho=0,z,t) \simeq e^{-ibz/V}
\int_{\om_{min}}^{\om_{max}}d\om \,
S(\om)\,e^{i\om\zeta/V}\,\left[H(z)-H\left(z-\frac{R}{\tan\theta(\om)}
\right)\right] \;, \label{tnp1} \ee

where $H(.)$ is the Heaviside step function and, let us recall,
$\theta(\om)$ is the axicon angle of the Bessel beam with angular
frequency $\om$.

\h We should notice that in the integrand of (\ref{tnp1}), the
step function $H(z-R/\tan\theta(\om))$ depends on $\om$, through
of $\theta(\om)$.

\h We can rewrite Eq.(\ref{tnp1}) in the form

\bb \begin{array}{clcr} \Psi_{TNP}(\rho=0,z>0,t) \simeq
&\dis{e^{-ibz/V}\left[ \int_{\om_{min}}^{\om_{max}}
S(\om)\,e^{i\om\zeta/V}\,d\om\right.}\\

\\

\\

&\dis{\left. - \int_{\om_{min}}^{\om_{max}}S(\om)\,e^{i\om\zeta/V}
\,H\left(z-\frac{R}{\tan\theta(\om)}
\right)d\om\right]}\end{array}\label{tnp} \ee

\

where the first term in the r.h.s. of (\ref{tnp}) is nothing but
the INP $\Psi_{INP}(\rho=0,z,t)$, while the second term is the
perturbation due to the truncation.

\h Now, remembering that for a Bessel beam of axicon angle
$\theta$ we have $\om = c\beta/\cos\theta$, and that the spectra
of INPs impose the constraint $\om=V\beta+b$ between angular
frequencies and longitudinal wavenumbers, it becomes easy to show
that

\bb \frac{R}{\tan\theta(\om)} \ug \frac{R}{\dis{\sqrt{1-\left(
\frac{c}{V} - \frac{bc}{V\om} \right)^2 }}} \left( \frac{c}{V} -
\frac{bc}{V\om} \right) \ee

and thus   

\bb \begin{array}{clcr} H\left(z-\frac{R}{\tan\theta(\om)}\right)
&\ug H\left(z-\frac{R\left( \frac{c}{V} - \frac{bc}{V\om}
\right)}{\sqrt{1-\left( \frac{c}{V} - \frac{bc}{V\om} \right)^2
}}\right)\\

\\&\ug \left\{\begin{array}{l} 1 \;\;\; {\rm for}\;\;\; z \geq
\frac{R\left( \frac{c}{V} - \frac{bc}{V\om}
\right)}{\sqrt{1-\left( \frac{c}{V} - \frac{bc}{V\om} \right)^2
}}\\

\\

0 \;\;\; {\rm for}\;\;\; z < \frac{R\left( \frac{c}{V} -
\frac{bc}{V\om} \right)}{\sqrt{1-\left( \frac{c}{V} -
\frac{bc}{V\om} \right)^2 }}\end{array}\right.\end{array}
\label{H} \ee

\

\h With all what precedes, we can eventually write

\

\bb \begin{array}{clcr} \Psi_{TNP}(\rho=0,z,t) \simeq
&\dis{e^{-ibz/V}\left[\int_{\om_{min}}^{\om_{max}}
S(\om)\,e^{i\om\zeta/V}\,d\om\right.}\\

\\

\\

&\dis{\left. - \int_{\om_{min}}^{\om_{max}}S(\om)\,e^{i\om\zeta/V}
\,H\left(z-\frac{R\left( \frac{c}{V} - \frac{bc}{V\om}
\right)}{\sqrt{1-\left( \frac{c}{V} - \frac{bc}{V\om} \right)^2
}}\right)d\om\right]}\end{array}\label{tnp2} \ee

\

\h In the next subsection, we will analyze the fundamental
Eq.(\ref{tnp2}) for the three possible types of TNPs: subluminal,
luminal, and superluminal.

\

\subsubsection{Subluminal TNP}

\h For the subluminal pulses ($V<c$), we have $b>0$, $\om_{min}=b$ and
$\om_{max} = cb/(c-V)$.

\h In this way, taking into account these facts, and that $z\geq0$
and $\om_{min} \leq \om \leq \om_{max}$, we can show, after
several manipulations, that Eq.(\ref{H}) can be written as

\bb H\left(z-\frac{R\left( \frac{c}{V} - \frac{bc}{V\om}
\right)}{\sqrt{1-\left( \frac{c}{V} - \frac{bc}{V\om} \right)^2
}}\right)\ug \left\{\begin{array}{clcr} 1 \;\;\; {\rm for}\;\;\;
\om \leq \dis{\frac{bc}{\dis{c-\frac{zV}{\sqrt{z^2+R^2}}}}}\\

\\

0 \;\;\; {\rm for}\;\;\; \om >
\dis{\frac{bc}{\dis{c-\frac{zV}{\sqrt{z^2+R^2}}}}}\end{array}\right.
 \label{Hsub} \ee

\
\

\h Now, by noting that $\om_{min}=b < bc/(c-zV/\sqrt{z^2+R^2}) <
\om_{max}=bc/(c-V)$, one can write Eq.(\ref{tnp2}), for the {\it
subluminal} case, as

\bb \begin{array}{clcr} \Psi_{TNP}(\rho=0,z>0,t) &\simeq
\dis{e^{-ibz/V}\left[\int_{b}^{\frac{bc}{c-V}}
S(\om)\,e^{i\om\zeta/V}\,d\om -
\int_{b}^{\frac{bc}{c-\frac{zV}{\sqrt{z^2+R^2}}}}
\, S(\om)\,e^{i\om\zeta/V}\,d\om\right]}    \\

\\

\\

&\ug
\dis{e^{-ibz/V}\left[\int_{\frac{bc}{c-\frac{zV}{\sqrt{z^2+R^2}}}}^{\frac{bc}{c-V}}
\, S(\om)\,e^{i\om\zeta/V}\,d\om\right]} \;\;\; ,\end{array}
\label{tnpsub} \ee

\

{\bf which represents our method in the case of subluminal} TNPs. \
It is a very simple equation, capable of providing closed-form, analytical
results for several different frequency spectra $S(\om)$, as we
shall see in Section 3.

\

\subsubsection{Luminal TNP}

\h For luminal TNPs ($V=c$), we have $b>0$, \ $\om_{min}=b$ and
$\om_{max} = \infty$.

\h With this, and taking into account that $z\geq0$ and $\om_{min}
\leq \om \leq \om_{max}$, we can, after several manipulations,
show that Eq.(\ref{H}) may be written as

\bb H\left(z-\frac{R\left( \frac{c}{V} - \frac{bc}{V\om}
\right)}{\sqrt{1-\left( \frac{c}{V} - \frac{bc}{V\om} \right)^2
}}\right)\ug \left\{\begin{array}{clcr} 1 \;\;\; {\rm for}\;\;\;
\om \leq \dis{\frac{b}{\dis{1-\frac{z}{\sqrt{z^2+R^2}}}}}\\

\\

0 \;\;\; {\rm for}\;\;\; \om >
\dis{\frac{b}{\dis{1-\frac{z}{\sqrt{z^2+R^2}}}}}\;\;\;
,\end{array}\right.
 \label{Hl} \ee

\

and, by noting that $\om_{min}=b < b/(1-z/\sqrt{z^2+R^2}) <
\om_{max}=\infty$, we can write Eq.(\ref{tnp2}), for the {\it
luminal} case, as

\bb \begin{array}{clcr} \Psi_{TNP}(\rho=0,z>0,t) &\simeq
\dis{e^{-ibz/c}\left[\int_{b}^{\infty}
S(\om)\,e^{i\om\zeta/c}\,d\om -
\int_{b}^{\frac{b}{1-\frac{z}{\sqrt{z^2+R^2}}}}
\, S(\om)\,e^{i\om\zeta/c}\,d\om\right]}    \\

\\

\\

&\ug
\dis{e^{-ibz/c}\left[\int_{\frac{b}{1-\frac{z}{\sqrt{z^2+R^2}}}}^{\infty}
\, S(\om)\,e^{i\om\zeta/c}\,d\om\right]} \;\;\; ,\end{array}
\label{tnpl}\ee

\h Equation (\ref{tnpl}), {\bf which represents our method in the
case of luminal} TNPs, is very simple too, and can provide closed-form,
analytical results for many different frequency spectra $S(\om)$,
as we shall see in Section 3.

\

\subsubsection{Superluminal TNP}

\h For superluminal TNPs ($V>c$), the value of $b$, in the the
spectral constraint (\ref{cond2}), can assume negative or
positive values, i.e., $-\infty \leq b \leq \infty$. Let us analyze the
superluminal case of Eqs.(\ref{H}) and (\ref{tnp2}) for both
situations, $b < 0$ and $b \geq 0$.

\

\h {\it\bf Superluminal case for $b < 0$}

\h In this case, we have $\om_{min}=cb/(c-V)$ and $\om_{max} =
\infty$.

\h Taking into account that $z\geq0$ and $\om_{min} \leq \om \leq
\om_{max}$, and, again, after several manipulations, we can show that for
this case Eq.(\ref{H}) can be written as

\

\bb H\left(z-\frac{R\left( \frac{c}{V} - \frac{bc}{V\om}
\right)}{\sqrt{1-\left( \frac{c}{V} - \frac{bc}{V\om} \right)^2
}}\right)\ug \left\{\begin{array}{l} 1 \;\;\; {\rm for}\;\;\;
\om \leq \dis{\frac{bc}{\dis{c-\frac{zV}{\sqrt{z^2+R^2}}}}} \;\;\; {\rm and}\;\;\; z \leq \frac{R}{\sqrt{V^2/c^2 -1}}\\

\\

1 \;\;\; {\rm for}\;\;\;
\om \geq \dis{\frac{bc}{\dis{c-\frac{zV}{\sqrt{z^2+R^2}}}}} \;\;\; {\rm and}\;\;\; z \geq \frac{R}{\sqrt{V^2/c^2 -1}}\\

\\

0 \;\;\; {\rm otherwise}\end{array}\right.
 \label{Hsup1} \ee

\

 \h By noting that, when $z \leq R/\sqrt{V^2/c^2 -1}$, we have
 $bc/(c-zV/\sqrt{z^2+R^2}) \leq
 0$ and that, when $z \geq R/\sqrt{V^2/c^2 -1}$, we have
 $bc/(c-zV/\sqrt{z^2+R^2}) > \om_{min} = bc/(c-V)$, one can write
 Eq.(\ref{tnp2}), for the case $V>c$ and $b<0$, as

\

\bb \Psi_{TNP}(\rho=0,z>0,t) \simeq \left\{ \begin{array}{clcr}
&\dis{e^{-ibz/V}\int_{\frac{bc}{c-V}}^{\infty}
S(\om)\,e^{i\om\zeta/V}\,d\om} \;\; &{\rm for}\;\;\; z \leq
\frac{R}{\sqrt{V^2/c^2 -1}}\\

\\

\\

\\

&\dis{e^{-ibz/V}\int_{\frac{bc}{c-V}}^{\frac{bc}{c-\frac{zV}{\sqrt{z^2+R^2}}}}
\,S(\om)\,e^{i\om\zeta/V}\,d\om } \;\; &{\rm for}\;\;\; z \geq
\frac{R}{\sqrt{V^2/c^2 -1}} \;\;\; ,\end{array}\right.
\label{tnpsup1}\ee

\

\h Equation (\ref{tnpl}), {\bf represents our method in the case
of superluminal} TNPs {\bf with} $b<0$. Again, the integrals are
very simple, and can provide closed-form, analytical results for many
spectra $S(\om)$.

\h Before going on to the next case, one can immediately see from
Eq.(\ref{tnpsup1}) that, independently of $S(\om)$, the
superluminal TNPs with $b<0$ will reach the distance
$z=R/\sqrt{V^2/c^2 -1}$ without deforming.

\

\h {\it\bf Superluminal case for $b \geq 0$}

\h In this case, we have $\om_{min}=b$ and $\om_{max} = \infty$.

\h Remebering that $z\geq0$ and $\om_{min} \leq \om \leq
\om_{max}$, and, as before, after several manipulations, we can show that
in this case Eq.(\ref{H}) can be written in the same form of
Eq.(\ref{Hsup1}).

\h Now, taking into account that, when $z \leq R/\sqrt{V^2/c^2
-1}$, it is $\om_{min}=b \leq bc/(c-zV/\sqrt{z^2+R^2}) \leq
\om_{max}=\infty $ and that, when $z > R/\sqrt{V^2/c^2 -1}$, one
has $bc/(c-zV/\sqrt{z^2+R^2}) < 0 $, we can write our fundamental
Eq.(\ref{tnp2}), for the case $V>c$ and $b \geq 0$, in the form

\

\bb \Psi_{TNP}(\rho=0,z>0,t) \simeq \left\{ \begin{array}{clcr}
&\dis{e^{-ibz/V}\int_{\frac{bc}{c-\frac{zV}{\sqrt{z^2+R^2}}}}^{\infty}
\,S(\om)\,e^{i\om\zeta/V}\,d\om} \;\; &{\rm for}\;\;\; z \leq
\frac{R}{\sqrt{V^2/c^2 -1}}\\

\\

&0 \;\; &{\rm for}\;\;\; z \geq \frac{R}{\sqrt{V^2/c^2 -1}} \;\;\;
,\end{array}\right. \label{tnpsup2}\ee

\

\h Equation (\ref{tnpl}), {\bf represents our method in the case
of superluminal} TNPs {\bf with} $b\geq 0$. Again, many closed-form
results, for many different spectra $S(\om)$, can be obtained from
the above equation.

\h We can also notice from Eq.(\ref{tnpsup2}) that, for $V>c$ and
$b \geq 0$, the superluminal TNPs will get very low intensities
after the distance $z=R/\sqrt{V^2/c^2 -1}$.

\h It is important to notice that in our method, given by
Eqs.(\ref{tnpsub}), (\ref{tnpl}), (\ref{tnpsup1}, \ref{tnpsup2}),
the on-axis evolution of a TNP depends only on the frequency
spectrum $S(\om)$ of its corresponding INP $\Psi_{INP}$, at variance with
the Rayleigh-Sommerfeld formula (\ref{rs}), which depends on the
mathematical expression of $\Psi_{INP}$.

\h Now, we shall go on to the next Section, where our method
will be applied to some important kinds of localized waves, and
our results will be compared with those obtained by making
numerical simulations by the Rayleigh-Sommerfeld formula.

\section{Application to some important cases. Closed-form, analytical
results, and their comparison with numerical simulations of the
Rayleigh-Sommerfeld formula.}

\h The method we have developed in the previous Section is
described by Eqs.(\ref{tnpsub}), (\ref{tnpl}), (\ref{tnpsup1},
\ref{tnpsup2}), for the cases of truncated subluminal, luminal and
superluminal pulses, respectively.

\h We are going to apply, now, this method to some important cases
involving TNPs; and it will be shown that the results agree very
well with those obtained through numerical simulations of the
Rayleigh-Sommerfeld formula (\ref{rs}).

\subsection{The method applied to a truncated subluminal pulse}

\h A well known ideal subluminal ($V<c$) nondiffracting pulse[20]
is

\bb \Psi_{INP}(\rho,z,t) \ug {\rm
exp}\left(\frac{i\,b\,V\,\gamma^2\,\eta}{c^2}\right)\,{\rm
sinc}\left(\frac{b\gamma}{c}\sqrt{\rho^2 + \gamma^2\zeta^2}\right)
\label{isub} \ee

\

where $\gamma = 1/\sqrt{1-V^2/c^2}$, $\eta=z-c^2t/V$ and, as
always, $\zeta=z-Vt$.

\h Now, we want to describe the on-axis behavior of the {\it
truncated} version ($\Psi_{TNP}$) of the the ideal solution
(\ref{isub}).

\h The subluminal INP (\ref{isub}) is generated by the superposition
(\ref{inp}), with a constant spectrum $S(\om)=c/(2bV\gamma^2)$.
However, we have to notice that this solution possesses backwards-travelling
components: Actually, it has $\om_{min}=bc/(c+V)$, instead of
$\om_{min}=b$ (which assures forward components only in
(\ref{inp}), as we have seen in Section 2). It is the price paid
to get such a closed-form and exact INP solution.

\h Anyway, we may, and we must, minimize the contribution of those
``backwards" components by choosing the subluminal velocity $V$ in
such a way that $(c+V)/(c-V)>>1$. Once this condition is satisfied, we
can observe that the INP (\ref{isub}) is, then, similar to that which
would be obtained with the same $S(\om)$, but with $\om_{min}=b$.

\h It should be noticed that the comments and observations above
have nothing to do with our method, which is constructed since the
beginning in order to comprehend causal (forward) solutions only, and
can be used for any values of the velocity $V$. Those remarks were
made just in order that a causal behavior of the INP (\ref{isub})
be guaranteed.

\h Now, on using $S(\om)=c/(2bV\gamma^2)$ in our Eq.(\ref{tnpsub}),
which describes the on-axis evolution of the subluminal TNPs, one
gets

\bb \begin{array}{clcr} \Psi_{TNP}(\rho=0,z>0,t) &\simeq
\dis{\frac{c}{2bV\gamma^2}}\,\dis{e^{-ibz/V}\left[\int_{\frac{bc}{c-\frac{zV}{\sqrt{z^2+R^2}}}}^{\frac{bc}{c-V}}
\,e^{i\om\zeta/V}\,d\om\right]} \\

\\

\\

&\ug \dis{\frac{cV}{2bV\gamma^2i\zeta}}e^{-ibz/V}\left[{\rm
exp}\left(\dis{\frac{ibc}{V(c-V)}\,\zeta}\right)- {\rm
exp}\left(\dis{\frac{ibc\sqrt{z^2+R^2}}{V(c\sqrt{z^2+R^2} -
zV)}\,\zeta}\right) \right]\; , \\
\\

\end{array} \label{tnpsub2} \ee

which is a very simple closed-form, analytical expression.

\h First, we use Eq.(\ref{tnpsub2}) to get the pulse-peak
intensity behavior. To do this, we just put $z=Vt \rightarrow
\zeta=0$ in (\ref{tnpsub2}).

\h Let us consider three different cases: (1) $V=0.995c\;$ and
$\;b=1.5\times 10^{15}$Hz; \   (2) $V=0.998c\;$ and $\;b=6\times
10^{14}$Hz, \ and (3) $V=0.9992c\;$ and $\;b=2.4\times 10^{14}$Hz.
In all cases, we consider the radius of the finite aperture to be
$R=4\;$mm.

\h At the same time, we compare our result with that obtained by the
numerical simulation of the Rayleigh-Sommerfeld formula
(\ref{rs}), with $\Psi_{INP}$ given by (\ref{isub}).

\h Figure 2 shows the plots, where those represented by continuous
lines have been obtained from our equation (\ref{tnpsub2}), while those
represented by dotted lines come from the numerical simulation of
(\ref{rs}).

\h We can verify the excellent agreement existing among those results.


\begin{figure}[!h]
\begin{center}
 \scalebox{3.4}{\includegraphics{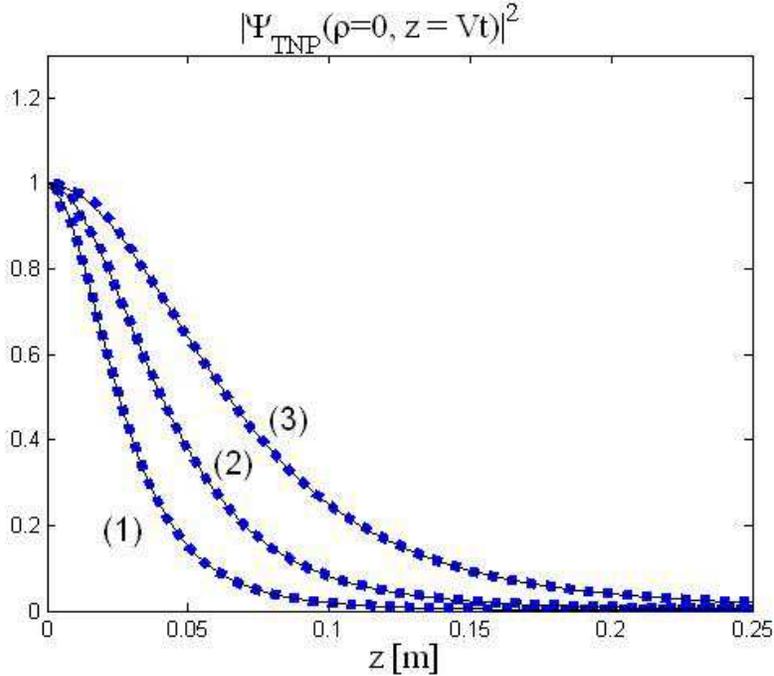}}
\end{center}
\caption{The peak intensity evolution of this subluminal TNP for
the three cases: (1) $V=0.995c\;$ and $\;b=1.5\times 10^{15}$Hz.
(2) $V=0.998c\;$ and $\;b=6\times 10^{14}$Hz. (3) $V=0.9992c\;$
and $\;b=2.4\times 10^{14}$Hz.\ In all cases $R=4\;$mm. The
continuous lines are obtained from our closed-form analytical
expression (\ref{tnpsub2}) and those represented by dotted lines
come from the numerical simulation of the Rayleigh-Sommerfeld
formula (\ref{rs}).} \label{Fig2}
\end{figure}

\

\h Now, we are interested in the on-axis longitudinal pulse
evolution, in the three cases above considered, in the time
instants given by $t'=0.11\;$ns, $t''=0.22\;$ns and
$t'''=0.33\;$ns.


\h Figures (3a,3b,3c) show the results corresponding to the cases
(1), (2) and (3), respectively. As before, the continuous lines
represent the results obtained from our Eq.(\ref{tnpsub2}), and
the dotted ones those coming from the numerical simulation of
(\ref{rs}). Again, we consider $R=4\;$mm.


\begin{figure}[!h]
\begin{center}
 \scalebox{1.75}{\includegraphics{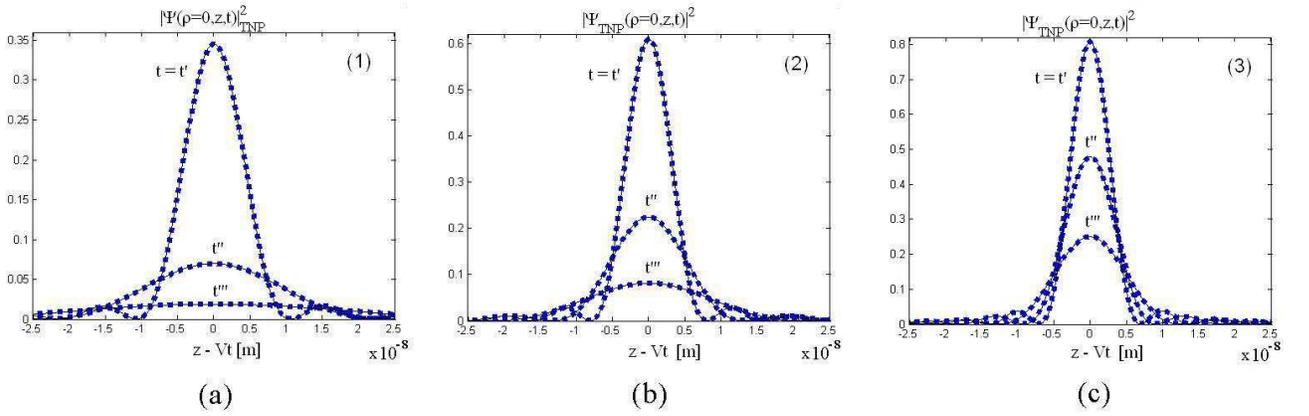}}
\end{center}
\caption{The on-axis evolution of this subluminal TNP, in the
times $t'=0.11$ns, $t''=0.22$ns and $t'''=0.33$ns, for each case
cited in Fig.2. \ Figure 3a, 3b and 3c represent the cases (1),(2)
and (3) respectively.\ The continuous lines are the results
obtained from our closed-form analytical expression (\ref{tnpsub2}) and
those represented by dotted lines come from the numerical
simulation of the Rayleigh-Sommerfeld formula (\ref{rs}).}
\label{Fig3}
\end{figure}

\h We can observe, once more, a very good agreement among the
results, confirming that our method works very well.

\

\subsection{The method applied to the truncated luminal Focus Wave Mode pulse}

\h A very famous ideal luminal ($V=c$) nondiffracting pulse[1] is the
Focus Wave Mode pulse (FWM), given by

\bb \Psi_{INP}(\rho,z,t) \ug \frac{ac}{ac-i\zeta}\,\,{\rm
exp}\left(\frac{-ib}{2c}\eta\right){\rm
exp}\left(\frac{-b\rho^2}{2c(ac-i\zeta)} \right)\; , \label{fwm}
\ee

where $\eta=z+ct$, \ $\zeta=z-ct$, and $a>0$ is a constant.

\h Like all the INPs, the FWM possesses infinite energy and must be
truncated (i.e., generated by a finite aperture) for real
applications. We shall use our method to get closed-form, analytical
expressions for the on-axis evolution of the truncated version
$\Psi_{TNP}$ of Eq.(\ref{fwm}).

\h The exact ideal solution (\ref{fwm}) is obtained from the
superposition (\ref{inp}) with $V=c$ and $S(\om)=a{\rm
exp}(ab/2){\rm exp}(-a\om)$. Because it has $\om_{min}=b/2$,
instead of $\om_{min}=b$, its spectrum possesses backwards
components[1-6] in the range $b/2 \leq \om < b$. To overcome this
problem, we must minimize the contribution of the nonphysical
part of the spectrum; and this can be done if $ab<<1$. Once such a
condition is obeyed, the INP (\ref{fwm}) can be considered similar
to the one that we would obtain with the same frequency spectrum of the FWM,
but with $\om_{min}=b$.

\h Again, let us notice that such remarks are made just to validate the
causality of the INP (\ref{fwm}), and have nothing to do with our own
method.

\h Now, by using $S(\om)=a{\rm exp}(ab/2){\rm exp}(-a\om)$ in our
Eq.(\ref{tnpl}), we get the on-axis evolution of the truncated
FWM:

\bb \begin{array}{clcr} \Psi_{TNP}(\rho=0,z>0,t) &\simeq
ae^{ab/2}\dis{e^{-ibz/c}\int_{\frac{b}{1-\frac{z}{\sqrt{z^2+R^2}}}}^{\infty}
\, e^{-a\om}\,e^{i\om\zeta/c}\,d\om} \\

\\

\\

&\ug \dis{\frac{ac}{ac-i\zeta}}e^{ab/2}e^{-ibz/c}{\rm
exp}\dis{\left(\frac{-b\sqrt{z^2+R^2}\,(ac-i\zeta)}{c(\sqrt{z^2+R^2}\,-z)}
\right)} \; ,\end{array} \label{tnpl2}\ee

\

which is a very simple closed-form, analytical expression.

\h First, let us put $\zeta=0$ in our Eq.(\ref{tnpl2}) in order to get the
pulse-peak intensity behavior.

\h We consider three different cases: (1) $a=1.6\times 10^{-16}$
and $b=5\times 10^{11}$Hz; \  (2) $a=1.25\times 10^{-16}$s and
$b=3\times 10^{11}$Hz; \ and (3) $a=1\times 10^{-16}$ and $b=2\times
10^{11}$Hz. In all cases we adopt the aperture radius $R=2\;$mm.

\begin{figure}[!h]
\begin{center}
 \scalebox{3.4}{\includegraphics{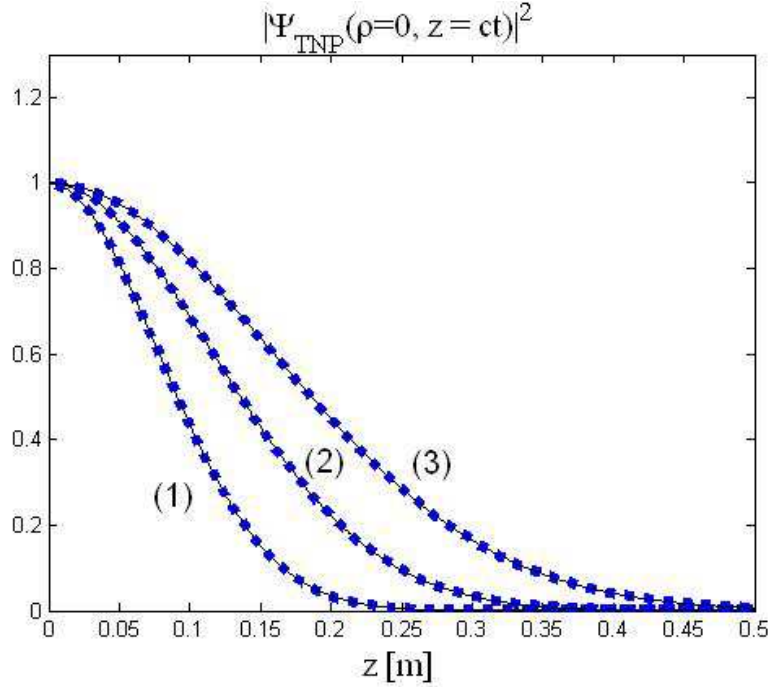}}
\end{center}
\caption{The peak-intensity evolution of the truncated luminal FWM
pulse for the three cases: (1) $a=1.6\times 10^{-16}$ and
$b=5\times 10^{11}$Hz; \ (2) $a=1.25\times 10^{-16}$s and
$b=3\times 10^{11}$Hz; \ (3) $a=1\times 10^{-16}$ and $b=2\times
10^{11}$Hz. \ In all cases $R=2\;$mm. The continuous lines are
obtained from our closed-form analytical expression (\ref{tnpl2}) and
those represented by dotted lines come from the numerical
simulation of the Rayleigh-Sommerfeld formula (\ref{rs}).}
\label{Fig4}
\end{figure}

\h Figure 4 shows the results. The continuous lines represent the
results obtained from our Eq.(\ref{tnpl2}), while the dotted ones are
the results of the numerical simulation of the Rayleigh-Sommerfeld
formula (\ref{rs}).

\h The results agree so well, that the corresponding continuous
and dotted curves do superpose to each other.

\h Now, we are going to use our Eq.(\ref{tnpl2}) to show the on-axix
evolution of this TNP, in the three cases above considered, for the time
instants $t'=0.22\;$ns, \ $t''=0.44\;$ns and $t'''=0.66\;$ns.

\h Figures (5a,5b,5c) show the results corresponding to the cases
(1), (2) and (3), respectively. The continuous lines are the results obtained
from our Eq.(\ref{tnpl2}), and the dotted lines those coming from
the numerical simulation of (\ref{rs}). Again, we consider
$R=2\;$mm.


\begin{figure}[!h]
\begin{center}
 \scalebox{1.75}{\includegraphics{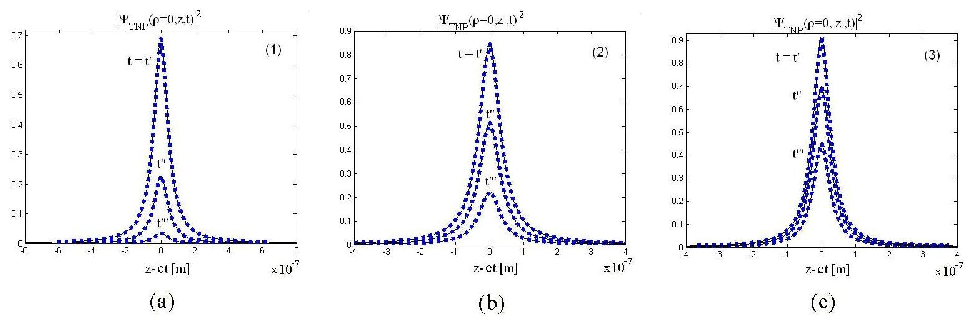}}
\end{center}
\caption{The on-axis evolution of the truncated luminal FWM pulse,
at the times $t'=0.22\;$ns, \ $t''=0.44\;$ns and $t'''=0.66\;$ns,
for each one of the cases cited in Fig.4. \ Figures 5a, 5b and 5c
represent the cases (1),(2) and (3), respectively.\ The continuous
lines are the results obtained from our closed-form analytical
expression (\ref{tnpl2}), while those represented by dotted lines
come from the numerical simulation of the Rayleigh-Sommerfeld
formula (\ref{rs}).} \label{Fig5}
\end{figure}

\h The results are in excellent agreement, showing the very good
efficiency of the method.


\subsection{The method applied to the truncated Superluminal Focus Wave Mode pulse}

\h An interesting, approximated, superluminal ($V>c$) ideal
nondiffracting solution to the wave equation is the so-called[6]
Superluminal Focus Wave Mode pulse (SFWM):

\bb \Psi_{INP}(\rho,z,t) \ug aV{\rm
exp}\left(\frac{-ib}{2V}\eta\right)\,X\,{\rm
exp}\left(\frac{b(V^2+c^2)}{2V(V^2-c^2)}\left((aV - i\zeta) -
X^{-1} \right) \right)\;, \label{sfwm} \ee

where $\eta=z+Vt$, \ $\zeta=z-Vt$, $a>0$ is a constant, and where

\bb X \ug \left((aV-i\zeta)^2 + \left(\frac{V^2}{c^2}-1
\right)\rho^2\right)^{-1/2} \label{X}\ee

\h Expression (\ref{tnpsup2}) is a very good aproximate
solution of the wave equation[6] if $ab<<1$, which is also the
condition for minimizing the contribution of the backwards components
of (\ref{tnpsup2}). \ Actually, this superluminal INP can be obtained
from superposition (\ref{inp}), with $b>0$, when using $S(\om)=a{\rm
exp}(ab/2){\rm exp}(-a\om)$, with constant $a>0$, but with
$\om_{min}=b/2$ instead of $\om_{min}=b$.

\h To get the closed-form, analytical mathematical expression that
describes the on-axis evolution of the truncated version of
(\ref{sfwm}), let us put $S(\om)=a{\rm exp}(ab/2){\rm exp}(-a\om)$ in
our Eq.(\ref{tnpsup2}):

\

\bb \Psi_{TNP}(\rho=0,z>0,t) \simeq \left\{ \begin{array}{clcr}
&\dis{\frac{aVe^{ab/2}e^{-ibz/V}}{aV-i\zeta}}\,\,{\rm
exp}\left(\dis{\frac{-bc\sqrt{z^2+R^2}\,(aV-i\zeta)}{V(c\sqrt{z^2+R^2}\,-
 zV)}} \right) &{\rm for}\;\; z \leq \frac{R}{\sqrt{V^2/c^2 -1}}  \\

\\

&0  &{\rm for}\;\; z \geq \frac{R}{\sqrt{V^2/c^2 -1}} \;\;\; ,\\

\\

\end{array}\right. \label{tnpsup21}\ee

\

\h Now, let us set $\zeta=0$ in our Eq.(\ref{tnpsup21}), for analyzing the
peak-intensity behavior of the truncated SFWM.

\h We consider three different cases: (1) $V=1.0002\;c$,
$b=3\times 10^{12}$Hz and $a=2.5\times 10^{-17}$s; \ (2)
$V=1.0001\;c$, $b=1\times 10^{12}$Hz and $a=5\times 10^{-17}$s; \
and (3) $V=1.00008\;c$, $b=2\times 10^{12}$Hz and $a=1.1\times
10^{-17}$s. In all these cases we choose $R=3\;$mm as being the
radius of the aperture.

\h The plots are shown in Fig.6, where the continuous lines
represent our results using Eq.(\ref{tnpsup21}), and the dotted
ones represent those obtained from the numerical simulation of
the Rayleigh-Sommerfeld formula (\ref{rs}). We can observe an
excellent agreement among the results.

\begin{figure}[!h]
\begin{center}
 \scalebox{3.4}{\includegraphics{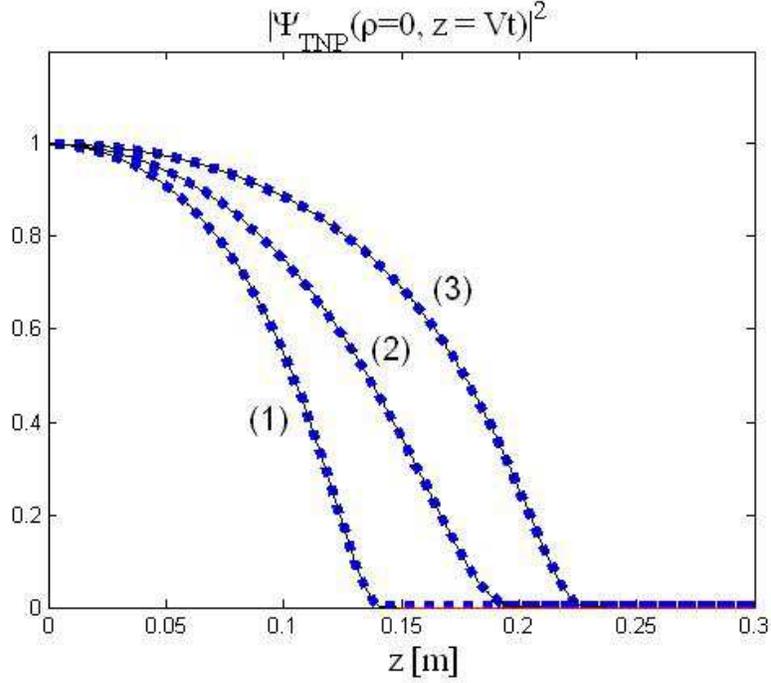}}
\end{center}
\caption{The peak intensity evolution of the truncated
superluminal Focus Wave Mode pulse for the three cases: (1)
$V=1.0002\;c$, $b=3\times 10^{12}$Hz and $a=2.5\times 10^{-17}$s;
\ (2) $V=1.0001\;c$, $b=1\times 10^{12}$Hz and $a=5\times
10^{-17}$s; \ and (3) $V=1.00008\;c$, $b=2\times 10^{12}$Hz and
$a=1.1\times 10^{-17}$s.\ In all cases $R=3\;$mm. The continuous
lines are obtained from our closed-form, analytical expression
(\ref{tnpsup21}), while those represented by dotted lines come
from the numerical simulation of the Rayleigh-Sommerfeld formula
(\ref{rs})} \label{Fig6}
\end{figure}

\h Now, we are going to use our method to show the on-axis
evolution of the pulse intensity at three different times
$t'=0.14\;$ns, $t''=0.29\;$ns and $t'''=0.43\;$ns, for each one of the
cases cited above.

\h Figures (7a,7b,7c) show the plots. Again, the curves given
by continuous lines come from our Eq.(\ref{tnpsup21}), while the
dotted ones from the numerical simulation of (\ref{rs}).

\begin{figure}[!h]
\begin{center}
 \scalebox{1.75}{\includegraphics{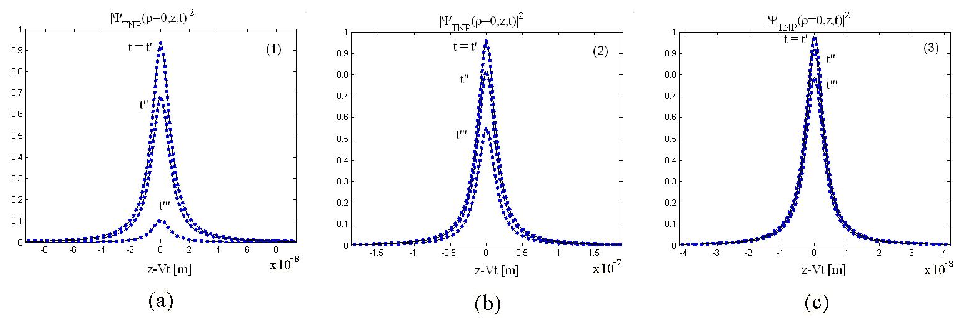}}
\end{center}
\caption{The on-axis evolution of the truncated SFWM pulse, at the
times $t'=0.14\;$ns, $t''=0.29\;$ns and $t'''=0.43\;$ns, for each
one of the cases cited in Fig.6. \ Figures 6a, 6b and 6c represent
the cases (1),(2) and (3), respectively.\ The continuous lines are
the results obtained from our closed-form, analytical expression
(\ref{tnpsup21}), and those represented by dotted lines come from
the numerical simulation of the Rayleigh-Sommerfeld formula
(\ref{rs}).} \label{}
\end{figure}

\h Once more, there is an excellent agreement among the results,
confirming the validity and efficiency of our method.


\

\section{Extending the present method to the almost nondiffracting
(finite energy) pulses, truncated by finite apertures}

\h In the previous Sections we have developed a (heuristic) method
capable of providing closed-form analytical expressions, describing the
on-axis evolution of the INP truncated by finite apertures.

\h It is well known[1,6] that, besides the INPs, there are the
almost nondiffracting pulses (ANP), which also need infinite
apertures to be generated, {\em but} possess a finite energy content.

\h Once a function $S(\om)$ is chosen, and an INP with a velocity $V$ and
$b=b_0$ is obtained from Eq.(\ref{inp}), we can get an ANP by
integrating Eq.(\ref{inp}), over the parameter
$b$, with a suitable choice of the weight function $S'(b)$ which has to be
concentrated around $b=b_0$. More explicitly:

\bb \Psi_{ANP}(\rho,z,t) \ug \int_{b_{min}}^{b_{max}} db\,
\int_{\om_{min}}^{\om_{max}}\,d\om\, S''(\om,b)
J_0\left(\rho\sqrt{\left(\frac{1}{c^2} - \frac{1}{V^2}
\right)\om^2 + \frac{2\,b\,\om}{V^2} -
\frac{b^2}{V^2}}\right)e^{i\om\zeta/V}e^{-ibz/V} \; , \label{anp} \ee

\

where $S''(\om,b)=S(\om)S'(b)$ is a spectral function with $S'(b)$
well localized around $b=b_0$.

\h Obviously, one can recover the INPs just by adopting the choice
$S''(\om,b)=S(\om)\delta(b-b_0)$.

\h An ANP can be viewed as a Bessel beam superposition
(\ref{bbsup}) with a spectral function $\overline{S}(\om,\beta)$
well concentrated around a straight line $\om = V\beta +
b_0$.

\h The ANPs are interesting solutions, due to their finite energy
contents, and can maintain their spatial shape for long
distances[1,6].

\h However, even possessing finite energy, the ANPs need ---as we
were saying--- infinite apertures to be generated: something that cannot
be obtained in the real world. Due to this fact, it is rather important
to know the behavior
of these pulses when truncated by finite apertures; that is, to
know the $\Psi_{TNPs}$ versions of the $\Psi_{ANPs}$. This can be
got by making numerical simulations, again, of the
Rayleigh-Sommerfeld integral formula (\ref{rs}), on replacing
$\Psi_{INP}$ with $\Psi_{ANP}$.

\h {\bf On the other hand, the extension of our method to the
cases of ANPs truncated by finite apertures can be performed in a very
simple way, just by multiplying our fundamental equations
(\ref{tnpsub}), (\ref{tnpl}) and (\ref{tnpsup1}, \ref{tnpsup2}) by
the $S(b')$ under consideration, and performing the relevant
integration over the parameter $b$.}

\h We are going to show this by the following example, in which we
shall obtain closed-form analytical expressions for the truncated version
$\Psi_{TNP}$ with a well known finite energy ANP[1,20]: Namely, the Modified
Power Spectrum pulse (MPS).

\

\h {\it Example}: {\large The truncated version of the MPS pulse.}


\h A well-known luminal ANP is the MPS pulse, which can be
obtained by integrating, over the parameter $b$, the
FWM pulse (\ref{fwm}) with the weight function

\bb S'(b) \ug H(b-b_0)\,q\,{\rm exp}(-q(b-b_0)) \; , \label{specmps}\ee

quantities $q$ and $b_0$ being positive constants. More explicitly, the
MPS pulse can be written as:

\bb \begin{array}{clcr} \Psi_{ANP}(\rho,z,t) &=
\dis{\int_{b_0}^{\infty}\frac{aqc}{ac-i\zeta}\,\,{\rm
exp}\left(\frac{-ib}{2c}\eta\right){\rm
exp}\left(\frac{-b\rho^2}{2c(ac-i\zeta)} \right){\rm
exp}(-q(b-b_0))}\,\,db \\

\\

&= \dis{\frac{2ac^2q}{(2cq+i\eta)(ac-i\zeta)+\rho^2}\,\,{\rm
exp}\left( \frac{-ib_0}{2c}\eta \right){\rm
exp}\left(-\frac{b_0\rho^2}{2c(ac-i\zeta)}\right)}\; , \end{array}
\label{mps} \ee

with $\eta=z+ct$, $\zeta=z-ct$.

\h This ANP has a finite energy content;  however, it needs an
infinite aperture to be generated.

\h We shall use the extended version of our method to get a
closed-form analytical expression for the on-axis evolution of the
truncated version, $\Psi_{TNP}$, of the MPS pulse. As we have
seen, to get this we just need multiplying the truncated version
of the FWM pulse, given by Eq.(\ref{tnpl2}), by the corresponding
weight function $S'(b)$ given by (\ref{specmps}), and performing
the integration over the parameter $b$. In this way, the on-axis
evolution of the truncated MPS pulse is given by:

\bb \begin{array}{l} \Psi_{TNP}(\rho=0,z>0,t) \simeq
\dis{\int_{b_0}^{\infty}\frac{aqc}{ac-i\zeta}e^{ab/2}e^{-ibz/c}{\rm
exp}(-q(b-b_0)){\rm
exp}\dis{\left(\frac{-b\sqrt{z^2+R^2}\,(ac-i\zeta)}{c(\sqrt{z^2+R^2}\,-z)}
\right)}}\,\,db \\

\\

\\

\;\;\;\;\;\;\;\;\;\;\;\;\;\;=\dis{
\frac{a\,q\,c\,e^{ab_0/2}e^{-ib_0z/c}}{\dis{(ac-i\zeta)\left(q -
\frac{a}{2} + \frac{iz}{c} +
\frac{\sqrt{z^2+R^2}\,(ac-i\zeta)}{c(\sqrt{z^2+R^2} - z )} \right)
}}\,\,{\rm
exp}\left(-\,\frac{b_0\sqrt{z^2+R^2}\,(ac-i\zeta)}{c(\sqrt{z^2+R^2}
- z )} \right)} \; ,\end{array} \label{tnpmps}\ee

\

which is a closed-form analytical expression.

\h As before, let us put $\zeta=0$ in our Eq.(\ref{tnpmps}) in order to get
the pulse peak intensity behavior.

\begin{figure}[!h]
\begin{center}
 \scalebox{3}{\includegraphics{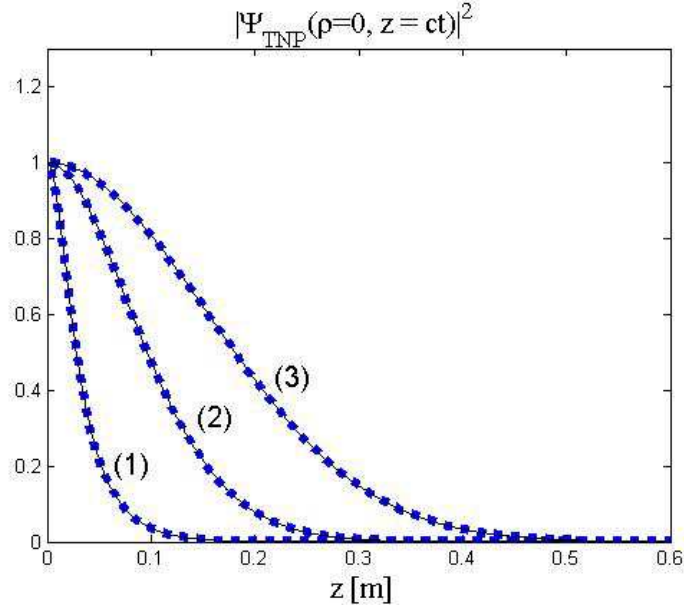}}
\end{center}
\caption{The peak-intensity evolution of the truncated luminal MPS
pulse for the three cases: (1) $a=1.6\times 10^{-16}$,
$b_0=5\times 10^{11}$Hz and $q=2 \times 10^{-11}$s; \ (2)
$a=1.25\times 10^{-16}$s, $b_0=3\times 10^{11}$Hz and $q=10 \times
10^{-11}$s; \ and (3) $a=1\times 10^{-16}$, $b_0=2\times
10^{11}$Hz and $q=20 \times 10^{-11}$s. \ In all cases $R=2\;$mm.
The continuous lines are obtained from our closed-form analytical
expression (\ref{tnpmps}), while those represented by dotted lines
come from the numerical simulation of the Rayleigh-Sommerfeld
formula (\ref{rs}).} \label{Fig8}
\end{figure}

\h Let us consider three different cases: (1) $a=1.6\times
10^{-16}\,$s, $b_0=5\times 10^{11}\,$Hz and $q=2 \times
10^{-11}\,$s; \ (2) $a=1.25\times 10^{-16}\,$s, $b_0=3\times
10^{11}\,$Hz and $q=10 \times 10^{-11}\,$s; \ and (3) $a=1\times
10^{-16}\,$s, $b_0=2\times 10^{11}\,$Hz and $q=20 \times
10^{-11}\,$s. In all cases, we adopt the aperture radius
$R=2\;$mm.

\h Figure 8 shows the results. The continuous lines represent
those obtained from our Eq.(\ref{tnpmps}), while the dotted ones
are the results of the numerical simulation of the
Rayleigh-Sommerfeld integral formula (\ref{rs}).

\h One can verify the excellent agreement among the results.

\h Now, we are going to use our Eq.(\ref{tnpmps}) to investigate the
on-axis evolution of this TNP in the three cases above considered,
for the instants $t'=0.22\;$ns, \ $t''=0.44\;$ns and
$t'''=0.66\;$ns.

\h Figures (9a,9b,9c) show the results corresponding to the cases
(1), (2) and (3), respectively. The continuous lines come from our
Eq.(\ref{tnpmps}), while the dotted lines are those coming from the
numerical simulation of (\ref{rs}). Again, we consider $R=2\;$mm.


\begin{figure}[!h]
\begin{center}
 \scalebox{1.75}{\includegraphics{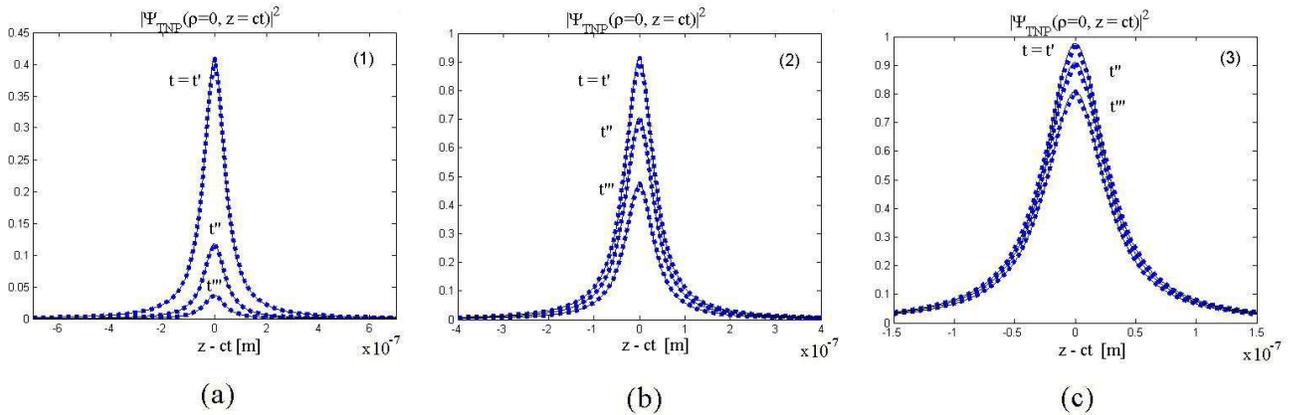}}
\end{center}
\caption{The on-axis evolution of the truncated luminal MPS pulse,
at the times $t'=0.22\;$ns, \ $t''=0.44\;$ns and $t'''=0.66\;$ns,
for each one of the three cases considered in Fig.8. \ Figures 9a, 9b and 9c
represent the cases (1),(2) and (3), respectively.\ The continuous
lines are the results obtained from our closed-form analytical
expression (\ref{tnpmps}), while those represented by dotted lines
come from the numerical simulation of the Rayleigh-Sommerfeld
formula (\ref{rs}).} \label{Fig9}
\end{figure}

\h Once more, there is an excellent agreement among the results,
confirming the validity and efficiency of our method.

\h Before finishing this Section, it is important to notice that the
closed-form analytical expressions of the truncated ANP,
obtained with our method, can be advantageously used for comparison purposes
with the corresponding non-truncated ANP: thus illustrating the effects
due to the truncation, and, for example, telling us till what
distance we can use the 3D solution of the ANP as a good approximation
to the corresponding 3D TNP.

\section{Conclusions}

\h In this paper we have developed a very simple method
for describing the space-time on-axis evolution of truncated
nondiffracting pulses, be they subluminal, luminal or
superluminal.

\h It is important to notice that in our method, given by
Eqs.(\ref{tnpsub}), (\ref{tnpl}), and (\ref{tnpsup1}, \ref{tnpsup2}),
the on-axis evolution of a TNP depends only on the frequency
spectrum $S(\om)$ of its corresponding INP $\Psi_{INP}$, contrary
to the Rayleigh-Sommerfeld formula (\ref{rs}) which depends on
the explicit mathematical expression of $\Psi_{INP}$. We also have extended
our method to describe the truncated versions of the ANPs.

\h Due to such a simplicity, we can obtain closed-form analytical
expressions, which describing the on-axis evolution of innumerable TNPs.
In this paper we have done that for the truncated versions of a
few, very well-known, localized waves: subluminal, luminal, or
superluminal.

\h We have compared our results with those obtained through the
numerical simulation of the Rayleigh-Sommerfeld integrals, and we have
observed an excellent agreement among them, confirming the
efficiency of our method.

\h The present approach can be very useful, because it furnishes,
in general, closed-form analytical expressions, avoiding the need of
time-consuming numerical simulations; and also because such
expressions provide a powerful tool for exploring several
important properties of the truncated localized pulses: as their
depth of field, the longitudinal pulse behavior, the decaying
rates, etc.

\

\section{Acknowledgements}

The author is very grateful to Erasmo Recami, Hugo E.
Hern\'andez-Figueroa and Claudio Conti for continuous discussions
and kind collaboration. Thanks are also due to Jane M. Madureira.

\h This work was supported by FAPESP (Brazil).

\h E-mail address: mzamboni@dmo.fee.unicamp.br


\

\

\section{References}

\

[1] I.M.Besieris, A.M.Shaarawi and R.W.Ziolkowski, ``A
bi-directional traveling plane wave representation of exact
solutions of the scalar wave equation", {\em J. Math. Phys.},
vol.30, pp.1254-1269, June 1989.\hfill\break

[2] J.-y.Lu and J.F.Greenleaf, ``Nondiffracting X-waves: Exact
solutions to free-space scalar wave equation and their finite
aperture realizations", {\em IEEE Trans. Ultrason. Ferroelectr.
Freq. Control}, vol.39, pp.19-31, Jan.1992.\hfill\break

[3] J.Fagerholm, A.T.Friberg, J.Huttunen, D.P.Morgan and
M.M.Salomaa, ``Angular-spectrum representation of nondiffracting X
waves", {\em Phys. Rev., E}, vol.54, pp.4347-4352,
Oct.1996.\hfill\break

[4] P.Saari and K.Reivelt, ``Evidence of X-shaped
propagation-invariant localized light waves", {\em Phys. Rev.
Lett.}, vol.79, pp.4135-4138, Nov.1997.\hfill\break

[5] E.Recami, ``On localized X-shaped Superluminal solutions to
Maxwell equations", {\em Physica A}, vol.252, pp.586-610,
Apr.1998.\hfill\break

[6] M.Zamboni-Rached, E.Recami and H.E.Hern\'andez F., ``New
localized Superluminal solutions to the wave equations with finite
total energies and arbitrary frequencies," {\em Eur. Phys. J., D},
vol.21, pp.217-228, Sept.2002.\hfill\break

[7] M.A.Porras, S.Trillo, C.Conti and P.Di Trapani, ``Paraxial
envelope X-waves," {\em Opt. Lett.}, vol.28, pp.1090-1092, July
2003.\hfill\break

[8] G. Nyitray and S. V. Kukhlevsky, ``Distortion-free tight
confinement and step-like decay of fs pulses in free space",
arXiv:physics/0310057 v1 13 Oct 2003.\hfill\break

[9] M. Zamboni-Rached, H.E. Hernandez-Figueroa, E. Recami "Chirped
Optical X-type Pulses," {\em J. Opt. Soc. Am. A}, Vol. 21, pp.
2455-2463 (2004).\hfill\break

[10] Goodman J W, ``Introduction to Fourier Optics", 1968 (New
York: McGraw-Hill).\hfill\break

[11] R.W.Ziolkowski, I.M.Besieris and A.M.Shaarawi, ``Aperture
realizations of exact solutions to homogeneous wave-equations",
{\em J. Opt. Soc. Am., A}, vol.10, pp.75-87, Jan.1993.\hfill\break

[12] J.-y.Lu and J.F.Greenleaf, ``Experimental verification of
nondiffracting X-waves", {\em IEEE Trans. Ultrason. Ferroelectr.
Freq. Control}, vol.39, pp.441-446, May 1992.\hfill\break

[13] J.-y.Lu, H.-h.Zou and J.F.Greenleaf, ``Biomedical ultrasound
beam forming", {\em Ultrasound in Medicine and Biology}, vol.20,
pp.403-428, 1994.\hfill\break

[14] M. Zamboni-Rached, A. Shaarawi, E. Recami, ``Focused X-Shaped
Pulses", Journal of Optical Society of America A, Vol. 21, pp.
1564-1574 (2004).\hfill\break

[15] J. Durnin, J. J. Miceli and J. H. Eberly, ``Diffraction-free
beams," {\em Phys. Rev. Lett.}, Vol. 58, pp. 1499-1501
(1987).\hfill\break

[16] M. Zamboni-Rached, ``Stationary optical wave fields with
arbitrary longitudinal shape by superposing equal frequency Bessel
beams: Frozen Waves", {\em Optics Express}, Vol. 12, pp.4001-4006
(2004).\hfill\break

[17] M. Zamboni-Rached, E. Recami, H. Figueroa, "Theory of Frozen
Waves: Modelling the Shape of Stationary Wave Fields", {\em J.
Opt. Soc. Am. A}, Vol. 22, pp. 2465-2475 (2005).\hfill\break

[18] M. Zamboni-Rached, ``Diffraction-Attenuation Resistant Beams
in Absorbing Media", arXiv:physics/0506067 v2 15 Jun 2005.
\hfill\break

[19] S. V. Kukhlevsky, M. Mechler, ``Diffraction-free
sub-wavelength beam optics at nanometer scale", {\em Opt. Comm.},
Vol. 231, pp. 35-43 (2004).\hfill\break

[20] R. Donnelly and R. W. Ziolkowski, ``Designing localized
waves", Proc. R. Soc. Lond. A, Vol.440, pp.541-565 (1993).

\end{document}